\documentstyle[12pt,fleqn,cite]{article}

\def\ab {(a)}
\def\ten{(10)}
\def\be{\begin{equation}}
\def\ee{\end{equation}}
\def\bea{\begin{eqnarray}}
\def\eea{\end{eqnarray}}

\begin{document}
\begin{titlepage}
\begin{center}
\hfill hep-th/9911090\\
\hfill CERN-TH/99-343\\
\vskip .2in

{\Large \bf Charged Macroscopic type II Strings and their Networks}
\vskip .5in

{\bf Alok Kumar$^{1,2}$\\
\vskip .1in
{\em 1. Theory Division, CERN,\\
CH-1211, Geneva 23, Switzerland}\\
{\em 2. Institute of Physics,\\
Bhubaneswar 751 005, INDIA}}
\end{center}

\begin{center} {\bf ABSTRACT}
\end{center}
\begin{quotation}\noindent
\baselineskip 10pt
We write down charged macroscopic string solutions in type II string theories,
compactified on torii, and  present an explicit solution of the spinor 
Killing equations to show that they preserve 1/2 of the  
type II supersymmetries. The S-duality symmetry of 
the type IIB string theory in ten-dimensions is used to write down the 
$SL(2,Z)$ multiplets of such strings and the
corresponding $1/2$ supersymmetry conditions. 
Finally we present examples of planar string networks,
using charged macroscopic $(p,q)$-strings. An interesting feature of some of 
these networks, which preserve 1/4 supersymmetry, 
is a required alignment among three parameters, namely 
the orientation of strings, a $U(1)$ phase associated with the 
maximal compact subgroup of 
$SL(2, Z)$, and an (angular) parameter associated with 
a solution generating transformation, which is responsible for   
creating charges and currents on the strings. 
\end{quotation}
\vskip .2in
CERN-TH/99-343\\
November 1999\\
\end{titlepage}
\vfill
\eject


\section{Introduction}
The study of macroscopic string 
solutions \cite{gibbon,sen92,sen95,waldram}
has been of importance in the last decade in several contexts, 
such as in black-hole physics\cite{malda},
strong/weak duality applications\cite{string-string,sen95,harv,
schwarz95} etc. In the context of duality, 
they have been crucial 
in establishing several such symmetries of string theories. 
Prominent among these are the $SL(2, Z)$ duality\cite{schwarz95} 
symmetries of 
the type IIB string theory in ten dimensions and a string-string 
duality between the type IIA compactification on K3 and heterotic string
compactification on $T^4$\cite{string-string,sen95,harv}. 
The support for the later conjecture involved the construcation of 
certain BPS solutions carrying (1-form) gauge field charges. Such solutions 
for $K3$ compactified type IIA theory were obtained using 
Charged Macrscopic String solutions of the heterotic strings with 
$1/2$ supersymmetry\cite{sen95} 
and then by mapping them to the type II strings.
In the former case, howerver, only a neutral string 
solution was needed, as type IIB in ten dimensions does not have any 
1-form gauge potential. A full duality multiplet of such 
neutral string solutions
and the corresponding duality covaraint string tensions were also 
obtained in \cite{schwarz95}. 
More recently, $SL(2,Z)$ as well as other
$U$-duality multiplets of neutral string solutions have been used 
in constructing their networks with 1/4\cite{sen97}, 
1/8\cite{bhatt} and lower supersymmetries
in type II theories. They are also expected to provide 
further confirmations of the duality conjectures.

The network solutions\cite{sen97,bhatt,schwarz,dasg,rey,aha,lee,
thorla,sudipt,subir,zwieb,bergmann,4d,rdevi}
of type II strings have also found their 
applications elsewhere, namely in providing nonperturbative 
symmetry enhancements in orientifold models to show their 
matching with the F-theory predictions\cite{zwieb}. 
In addition, strings and networks
which end on D3-branes have found a wide application in 4-dimensional 
gauge\cite{bergmann,callan,4d} and other world-volume 
theories\cite{bhatt2}. 

The focus of attention in this paper are the Charged Macroscopic
String solutions\cite{sen92,sen95} 
and their networks. As stated earlier, these solutions
have been used earlier for constructing the soliton multiplets in type
II string compactification on K3 in order to provide support for their duality
conjecture with the heterotic theory on $T^4$. 
In this paper, howerver, we will concentrate on such solutions
in type II theories, when they are compactified on torii. As a result, 
a verification of supersymmetry requires an analysis of additional 
Killing equations than the ones which are present in the heterotic 
strings, namely one has to examine the supersymmetry conditions for 
the spinors arising from both the left and the right-moving sectors
of the type II theories. In this paper we perform this 
analysis explicitly for a class of such Charged Macroscopic 
String solutions which are analogous to the heterotic solutions 
presented in \cite{sen92}. We also write down explicit supersymmetry
conditions for several other class of examples in \cite{sen95}.

The Charged Macroscopic String solutions are generated from the 
neutral ones by a solution generating transformation
and are in general parameterized 
by a group $O(d-1, 1; d-1, 1)$, 
arising out of one time and  
$d-1$ spatial translational isometries 
of the solution. These parameters also appear 
in the Charged Macroscopic String solutions. In particular, 
the solutions in \cite{sen95} are characterized by two nontrivial 
$O(d-1, 1; d-1, 1)$ parameters $\alpha$ and $\beta$, which apply boost between 
the time direction and an internal direction in the left and the
right-moving sectors respectively. The solutions of \cite{sen92}, which we
use to explicitly show the $1/2$ supersymmetric nature of these solutions
in section-2
correspond to $\beta =0$, but $\alpha \neq 0$. 
Our analysis then suggests that
general solutions ($\alpha \neq 0, \beta \neq 0$)
also preserve $1/2$ supersymmetry.

Our results  show that $O(d-1, 1; d-1, 1)$ transformations parameterized by 
$\alpha$ and $\beta$ change the Killing equations in a 
nontrivial way. As a result, the supersymmetry conditions and the 
form of the Killing spinors is also modified. However 
both the supersymmetry
conditions and the Killing spinors for the charged string
can be generated from those for the
neutral ones by  Lorentzian tranformations. The parameters of these
Lorentz transformations turn out to be local, 
having a coordinate dependence on the transverse radius. 
The experience gained from this analysis  
(for $\beta=0$ ) can in fact 
be used to write down the supersymmetry condition for  other 
solutions characterized by parameters $\alpha$ and $\beta$. 
In view of our future application, in section-3 
we confirm the $1/2$ supersymmetry property of $\alpha = \beta$
and $\alpha = -\beta$ solutions by examining the consistency of 
the dilatino supersymmetry variation for the 
charged macroscopic string background. 
We also show that one of the above solutions, 
namely $\alpha = -\beta$, when decompactified to ten 
dimensions, is related to the neutral string solutions
by a constant coordinate transformation. This is not surprising, 
as $O(d,d)$ group is known to contain a $GL(d)$ 
subgroup of constant coordinate transformations. The other 
possibility, namely $\alpha = \beta$ that we have analyzed is
an inequivalent solution even in ten-dimensional sense. 
This can be verified from the 
expression for the dilaton, which is now different from the one
for the neutral string. However we like to point out that even 
$\alpha = -\beta$ solutions
are in fact physically different in the compactified 
theory and represent genuine charged 
strings in $D\leq 9$.

We then use the $SL(2,Z)$ duality symmetries of the type IIB theories in 
ten dimensions to generate general $(p,q)$-charged macroscopic string
solutions from the $(1,0)$ or elementary-string solution discussed above. 
In particular we show that the supersymmetry conditions for both 
$\alpha = \pm \beta \neq 0$ soultions are of a form which allow
the constructon of string networks preserving $1/4$ supersymmetry. 
This is not surprising for the $\alpha = -\beta$ solution for the reason 
already stated in the last paragraph. 
As a result, the $1/4$ 
supersymmetry of these networks already follows from that of the 
netutral planar string networks that exist in various dimensions. 
In this case, we find that the internal torii do not play any 
significant role and the string networks can be constructed by 
aligning the orinetation of the $(p,q)$-string, in a plane, with 
respect to a phase associated with the transformation of
spinors under the $SL(2,Z)$ duality symmetry transformation.

The charged string solutions with $\alpha =\beta$ turn 
out to be more interesting from the 
supersymmetry point of view for the construction of networks.
We find that in this case a network construction, preserving 
certain supersymmetry, requires not only an alignment between 
the two angles discussed above, but in 
addition, one has to further align them with an angle coming from 
the soultion generating parameter.  
In our examples, in section (4.3), these strings carry not only 
the 2-form charges parameterized by integers $(p,q)$ and 
moduli $\tau$, but also by gauge charges characterized by a 2-dimensional 
unit vector $\hat{n}$. Physically, this alignment therefore implies
a coupling between the $SL(2)$ charges with 
that of the gauge charges and also a relationship between 
their conservation laws.

The outline of the paper is as following. In section-2 we 
write down the general charged string solution in arbitrary dimensions.
Then to work out the supersymmetry, we restrict to a specific case,
namely $\beta=0$ and present the Killing spinors for this example. 
Although our analysis is performed specifically in 9-dimensions,
we present the generalizations of the results to other lower 
dimensions as well. In section-3 of the paper, we write down 
explicit supersymmetry conditions 
for $\alpha = \beta \neq 0$ and $\alpha = - \beta \neq 0$. 
Once again we show that our background fields satisfy a 
nontrivial condition required for the consistency of these
spinor equations. 
Again  the derivations are given explicitly 
in 9-dimensions and then generalized to the lower ones.
In section-4, following \cite{schwarz95}, we also
write the $SL(2,Z)$ multiplets of the charged macroscopic strings 
and show the existence of network solutions for the 
examples worked out in section-3.  
This is done by demonstrating the existence of 
a unique spinor at aymptotic infinity, satisfying the supersymmetry 
conditions for arbitrary number of $(p,q)$-strings, provided the 
alignments we referred previously, also hold. Discussions and 
conclusions are presented in section-5.

\section{Killing Spinors for a Charged Macroscopic String in 
$D\leq 9$}

{\bf (2.1) Bosonic Backgrounds}
\vskip 1cm

We start by writing down the bosonic backgrounds associated with 
the Charged Macroscopic strings in space-time dimensions $D$. 
They have been obtained from 
similar solutions for the heterotic strings\cite{sen95}, 
by turning off
the sixteen gauge fields associated with the right-moving,
bosonic sector. This is possible since this sector of the 
heterotic string is identical to the NS-NS sector of type II 
theories in ten dimensions. The solution is given by,
\bea \label{e13}
ds^2 &=& r^{D-4} \Delta^{-1} [ -(r^{D-4}+C) dt^2 + C (\cosh\alpha -
\cosh\beta) dt dx^{D-1}  \nonumber \\
&& + (r^{D-4} + C \cosh\alpha \cosh\beta) (dx^{D-1})^2]
\nonumber \\
&& + (dr^2 + r^2 d\Omega_{D-3}^2) \, ,
\eea
\be \label{e14}
B_{ (D-1)t} = {C\over 2\Delta} (\cosh\alpha + \cosh\beta)
\{ r^{D-4} + {1\over 2} C (1 + \cosh\alpha \cosh \beta) \} \, ,
\ee
\be \label {e15}
e^{-\Phi} = {\Delta^{1/2} \over r^{D-4}} \, ,
\ee
\bea \label{e16}
A^{\ab}_{ t} &=& -{n^{\ab} \over 2 \sqrt 2 \Delta} C \sinh \alpha
\{ r^{D-4} \cosh\beta + {1\over 2} C (\cosh\alpha + \cosh \beta) \}
\nonumber \\
&& \qquad \qquad \hbox{for} \qquad 1 \le a \le (10-D) \, , \nonumber \\
&=& -{p^{(a-10+D)} \over 2 \sqrt 2 \Delta} C \sinh \beta
\{ r^{D-4} \cosh\alpha + {1\over 2} C (\cosh\alpha + \cosh \beta) \}
\nonumber \\
&& \qquad \qquad \hbox{for} \qquad (10-D)+1 \le a \le (20 - 2D) \, , 
\nonumber \\
\eea
\bea \label{e17}
A^{\ab}_{ D-1} &=& - {n^{\ab} \over 2\sqrt 2 \Delta} C \sinh \alpha 
\{r^{D-4}
+ {1\over 2} C \cosh \beta (\cosh\alpha + \cosh \beta) \} \nonumber \\
&& \qquad \qquad \hbox{for} \qquad
1\le a \le (10-D) \, , \nonumber \\
&=&  {p^{(a-10+D)} \over 2\sqrt 2 \Delta} C \sinh \beta \{ r^{D-4}
+ {1\over 2} C \cosh \alpha (\cosh\alpha + \cosh \beta) \} \nonumber \\
&& \qquad  \qquad \hbox{for} \qquad
(10-D)+1 \le a \le (20-2D) \, , \nonumber \\
\eea
\be \label{e18}
M_D = I_{20-2D} + \pmatrix{ P nn^T & Q np^T \cr Q pn^T & P pp^T \cr} \, ,
\ee
where,
\be \label{e19}
\Delta = r^{2(D-4)} + Cr^{D-4} ( 1 + \cosh\alpha \cosh\beta) + {C^2 \over 4}
(\cosh\alpha + \cosh\beta)^2 \, ,
\ee
\be \label{e19a}
P = {C^2 \over 2\Delta} \sinh^2 \alpha \sinh^2 \beta \, ,
\ee
\be \label{e19b}
Q = - C \Delta^{-1} \sinh\alpha \sinh\beta \{ r^{D-4} + {1\over 2} C
(1 + \cosh\alpha \cosh \beta) \} \, .
\ee
with $n^{(a)}$, $p^{(a)}$ being the components of $(10-D)$-dimensional 
unit vectors. 
$A_{\mu}$'s in eqns. (\ref{e16}), (\ref{e17})
are the gauge fields appearing due to the Kaluza-Klein (KK)
reductions of the ten dimensional metric and the 2-form antisymmetric
tensor coming from the NS-NS sector. The matrix $M_D$ parametrizes
the moduli fields. The exact form of this parametrization 
depends on the form of the $O(10-D, 10-D)$ metric used. 
The above solution has been written for a diagonal
metric of the form: 
\be
 L_D = \pmatrix{ -I_{10-D} & \cr & I_{10-D}}. \label{eld}
\ee

Later on, while decompactifying these backgrounds, in order to 
check supersymmetry, we will use the notations and conventions in 
\cite{senijmp} which uses a different form of the 
metric, namely:
\be
L = \pmatrix{ & I_{10-D} \cr I_{10-D} & }. \label{el}
\ee
These two conventions are howerver related by:
\be
 L_D = \hat{P} L \hat{P}^T, \>\>\> M_D = \hat{P} M \hat{P}^T, \label{eleld}
\ee
where 
\be
   \hat{P} = {1\over \sqrt{2}}\pmatrix{-I_{10-D} & I_{10-D} \cr 
I_{10-D} & I_{10-D}}. \label{hatp}
\ee 

The gauge fields in two conventions are related as:
\be
  \pmatrix{A^1_{\mu} \cr A^2_{\mu}} = 
\hat{P} \pmatrix{\hat{A}^1_{\mu} \cr \hat{A}^2_{\mu}}, 
            \label{aredef}
\ee
with $A^{1,2}_{\mu}$'s in the above equation
being $(10-D)$-dimensional columns consisting 
of the gauge fields $A_{\mu}$'s defined in (\ref{e16}-\ref{e17}), 
and coming from the left and the right-moving sectors.  

In this section we now restrict ourselves to the $\beta=0$ 
solutions. These solutions are analogous to the ones written  
for the hetrotic strings in \cite{sen92} and are given by,
\bea
ds^{2} = {1\over \cosh^2{\alpha\over 2} e^{-E} -\sinh^2{\alpha\over
2}} (-dt^2 +(dx^{D-1})^2) \cr
& \cr
+{\sinh^2{\alpha\over 2} (e^{-E}-1)\over
(\cosh^2{\alpha\over 2} e^{-E} -\sinh^2{\alpha\over
2})^2} (dt+dx^{D-1})^2
+\sum_{i=1}^{D-2} dx^i dx^i\cr\cr
& \cr
B_{(D-1)t} = {\cosh^2{\alpha\over 2}(e^{-E}-1)\over \cosh^2{\alpha\over
2} e^{-E} -\sinh^2{\alpha\over
2}}\cr\cr
& \cr
A^{(1)}_{D-1} = A^{(1)}_t = - {1\over {2\sqrt{2}}}\times
{\sinh\alpha (e^{-E}-1)\over
\cosh^2{\alpha\over 2} e^{-E} -\sinh^2{\alpha\over
2}}.\cr 
& \cr
\Phi = -\ln(\cosh^2{\alpha\over 2}e^{-E} -\sinh^2{\alpha\over 2})
\label{betaeq0}
\eea
with $e^{-E}$ being the Green function in the $D-2$ dimensional 
transverse space:
\be
e^{-E} = (1 + {C\over r^{D-2}}). \label{green}
\ee
and constant $C$ determining the string tension. {\footnote{ 
There is an extra 
factor of ${1\over {2\sqrt{2}}}$ appearing in (\ref{betaeq0})
with respect to the one in \cite{sen92}. This howerver has been 
taken care in \cite{sen92} in the definitions of charges.}}

Now, in order to understand the 
type II origin of various background fields and to verify the 
supersymmetry of these solutions, we decompactify the above solution 
back to ten dimensions. 
The decompactification exercise is done following a set of notations 
given in \cite{senijmp}. When restricted to the NS-NS sector of 
type II theories, they can be written as:
\bea\label{decomp}
&& \hat {G}_{a b}  = G^{\ten}_{[a+(D-1), b+(D-1)]}, 
\quad  \hat B_{a b}  =
B^{\ten}_{[a+(D-1), b+(D-1)]}, 
\nonumber \\
&& \hat{A}^{(a)}_{\bar{\mu}}  = {1\over 2}\hat G^{ab} 
G^{\ten}_{[b+(D-1),\bar{\mu}]}, \nonumber \\
&&  \hat{A}^{(a+(10-D))}_{\bar{\mu}} = {1\over 2}
B^{\ten}_{[a+(D-1), \bar{\mu}]} - \hat B_{ab} A^{(b)}_{\bar{\mu}}, 
\nonumber \\
&& G_{\bar{\mu}\bar{\nu}} = G^{\ten}_{\bar{\mu}\bar{\nu}} 
- G^{\ten}_{[(a+(D-1)), \bar{\mu}]} 
G^{\ten}_{[(b+(D-1)), \bar{\nu}]} \hat
G^{ab}, \nonumber \\
&& B_{\bar{\mu}\bar{\nu}} = 
B^{\ten}_{\bar{\mu}\bar{\nu}} - 4\hat B_{ab} A^{(a)}_{\bar{\mu}}
A^{(b)}_{\bar{\nu}} - 
2 (A^{(a)}_{\bar{\mu}} A^{(a+(10-D))}_{\bar{\nu}} - A^{(a)}_{\bar{\nu}} 
A^{(a+(10-D))}_{\bar{\mu}}),
\nonumber \\
&& \Phi = \Phi^{\ten} - {1\over 2} \ln\det \hat G, \quad \quad
\quad 1\le a, b \le 10-D, \quad
0\le {\bar{\mu}}, \bar{\nu} \le (D-1).
\eea

We now start with a nine-dimensional ($D=9$) solution in 
(\ref{betaeq0}) and 
following the Kaluza-Klein (KK) compactification mechanism summarized above,
write down the solution directly in ten dimensions. 
We do this first for the $D=9$ solution and
later in section-(2.4) generalize the results to $D < 9$. 
Only nonzero background fields are then given by
\bea \label{10dg}
ds^2 =  {1\over {cosh^2 {\alpha\over 2} e^{-E} - sinh^2 {\alpha\over 2}}}
         ( -dt^2 + (dx^8)^2 ) +  
        {sinh^2{\alpha\over 2}(e^{-E} - 1)\over 
        {cosh^2 {\alpha\over 2}e^{-E} - sinh^2 {\alpha\over 2}}}
        (dt + dx^{8})^2 +  \cr 
& \cr
        + {{sinh \alpha (e^{-E} - 1)}\over 
                  {cosh^2 {\alpha\over 2}e^{-E} - sinh^2 {\alpha\over 2}}}
         dx^9 (dt + dx^8) +  \sum_{i=1}^{7} dx^i dx^i + (dx^9)^2,
\eea
\be \label{10db}
    B_{8 t}  = {{cosh^2 {\alpha\over 2} (e^{-E} - 1)}
              \over {cosh^2 {\alpha\over 2}e^{-E} - sinh^2 {\alpha\over 2}}},\>
\>\>
    B_{9 t}  = - {sinh\alpha\over 2}{{(e^{-E} - 1)}\over 
                  {cosh^2 {\alpha\over 2}e^{-E} - sinh^2 {\alpha\over 2}}} 
\>\>\>= B_{98}.
\ee
The dilaton in ten dimensions remains same as the one in 
(\ref{betaeq0}):
\be \label{10phi} 
   \phi^{(10)} = - ln ({cosh^2 {\alpha\over 2} e^{-E}
- sinh^2 {\alpha\over 2}}).
\ee
Although it is already expected, we have also reconfirmed that 
many of the field equations in ten-dimensions are satisfied by 
the backgrounds in eqns. (\ref{10dg}), (\ref{10db}), (\ref{10phi}).

We now study spinor Killing equatoins for type IIB strings in ten dimensions
and show that the solutions in (\ref{10dg})-(\ref{10phi}) 
are consistent with $1/2$ 
supersymmetry. We once again emphasize that $1/2$ supersymmetry 
from the type IIB string point of view is 
comparatively more nontrivial,
than in the heterotic theory, due 
to the presence of extra equations to be satisfied by the 
background configuration. Later in section-(2.3) we also 
find the corresponding Killing spinors. 
\vskip 1cm

{\bf (2.2) Killing Equations}
\vskip .7cm

The spinor Killing equations in ten dimensions, 
when restricted to NS-NS fields,  
follow from supersymmetry variations, 
(in string metric)\cite{waldram,schwarz83,hassan}: 
\be
\delta \psi_M  = \partial_M \eta + {1\over 4}
             \omega_M^{\hat{M} \hat{N}} 
             \Gamma_{\hat{M} \hat{N}} \eta - 
           {1\over 8} H_M^{\hat{M} \hat{N}} 
           \Gamma_{\hat{M} \hat{N}} \eta^*, 
           \label{gravitino}
\ee
\be
\delta \lambda  = (\partial_M \phi^{(10)}) \gamma^{M} \eta^* -
         {1\over 6} H_{M N P}\gamma^{M N P} \eta,  
                     \label{dilatino}
\ee
where $\psi_M$ is the ten-dimensional gravitino, $\lambda$ the 
dilatino and $\eta \equiv (\epsilon_L + i \epsilon_R)$ are the 
supersymmetry parameters. $M = 0,..,9$ are the general coordinate
indices in ten dimensions and 
$\hat{M}, \hat{N}$ are the Lorentz indices. 

To analyze these equations for our nine-dimensional 
solution, we now denote the indices {$(9,0,8)$}
by greek indices ${\mu}$. The corresponding Lorentz indices 
are denoted by $\hat{\mu}$ etc.. The indices, transverse to 
the string are denoted by $m = 1,..,7$ and the corresponding 
Lorentz ones by $\hat{m}$'s etc.. The 
ten-dimensional Lorentzian metric for our purpose is taken to 
be of the form: 
$\eta_{\hat{M} \hat{N}} \equiv diag.(1, -1, 1, ...,1)$ (with the first 
entry denoting the coordinate $x^9$), 
which implies: $\eta_{\hat{\mu} \hat{\nu}} = (1, -1, 1)$ and  
also $\eta_{\hat{m} \hat{n}} = \delta_{\hat{m} \hat{n}}$. 
Taking into account that the backgrounds depend only on 
transverse coordinates  denoted by $m$'s through radius $r$,
the gravitino supersymmetry variation (\ref{gravitino}) can 
be written as:
\bea
\delta \psi_m & = \partial_m \eta + {1\over 4}
\omega_m^{\hat{\mu} \hat{\nu}}\Gamma_{\hat{\mu} \hat{\nu}}\eta 
- {1\over 8} H_m^{\hat{\mu}\hat{\nu}}\Gamma_{\hat{\mu}\hat{\nu}}\eta^*, 
         \label{transv}\\
& \cr
\delta \psi_{\mu} & = {1\over 2} \omega_{\mu}^{\hat{\nu} \hat{m}}
                     \Gamma_{\hat{\nu} \hat{m}} \eta 
                     - {1\over 4} H_{\mu}^{\hat{\nu}\hat{m}}
                      \Gamma_{\hat{\nu}\hat{m}}\eta^* . 
                  \label{longi}
\eea

For the purpose of algebraic manipulations, 
we find it convenient to write these equations by
introducing parameters:
\bea
 g & =  {1\over {cosh^2 {\alpha\over 2}e^{-E} 
- sinh^2 {\alpha\over 2}}}, \>\>\>\cr
& \cr
 a & =  {sinh^2{\alpha\over 2}(e^{-E} - 1)\over 
        {(cosh^2 {\alpha\over 2}e^{-E} 
- sinh^2 {\alpha\over 2} )}^2} , \>\>\> \cr
& \cr
 b & =  {sinh \alpha\over 2} {{(e^{-E} - 1)}\over 
                   {cosh^2 {\alpha\over 2}e^{-E} - sinh^2 {\alpha\over 2}}},
                    \label{gab}
\eea
and $3\times 3$ matrices ${\cal G}_{\mu \nu}$, 
${\cal B}_{\mu \nu}$ and 
${\cal E}_{\mu}^{\hat{\mu}}$, where the metric ${\cal G}$ 
and the antisymmetric 
tensor ${\cal B}$ can be read from the backgrounds in 
eqns.(\ref{10dg})-(\ref{10phi}). 
${\cal E}$ is the vielbein correponding to ${\cal G}$. In our case
these $3\times 3$ matrices 
can be written in terms of $2\times 2$ matrices 
$G$, $B$ and $\hat{E}$:
\bea
    \cal{G}  = \pmatrix{1 & {\hat{b}} \cr {\hat{b}}^T 
& G + {\hat{b}}^T {\hat{b}}}
\>\>\> 
    \cal{B} = \pmatrix{0 & -{\hat{b}} \cr {\hat{b}}^T & B}
          \label{calbg}
\eea
with ${\hat{b}} \equiv b (1, 1)$,   
a 2-dimensional row-vector. The vielbein ${\cal E}$ is given by:
\bea
    \cal{E} = \pmatrix{1 & 0 \cr {\hat{b}}^T & {\hat{E}}}
             \label{calee}
\eea
and satisfies ${\cal E} \eta {\cal E}^T = {\cal G}$,
whereas 
$\hat{E} \hat{\eta} \hat{E}^T = G$, with 
$\hat{\eta}$ being a diagnoal $2\times 2$ matrix: $diag.(-1, 1)$.

The $2\times 2$ matrices $G$, $B$ and $\hat{E}$ apprearing 
in eqns. (\ref{calbg}), (\ref{calee}) have explicit forms:
\bea 
 G \equiv \pmatrix{ -g + a & a \cr a & g + a}, \label{9dg}
\eea
\bea
 B \equiv \pmatrix{ 0 & g-1 \cr 1 - g & 0 }, \label{9db}
\eea 
and
\bea
\hat{E} \equiv {1\over \sqrt{g-a}}\pmatrix{ g-a & 0 \cr
                                       -a & g }. \label{9dviel}
\eea
Note that $G$ also represents the longitudinal part, or (0,8)-components,
of the compactified metric in $D$-dimensions, as seen directly from 
eqn.(\ref{betaeq0}). Similarly $B$ is the antisymmetric tensor in the 
compactified theory and $\hat{E}$ is the vielbein for the 
metric $G$. Using these notations we now start by 
simplifying the gravitino variation equation for the transverse 
coordinates, $m$, namely eqn.(\ref{transv}). 

The spin-connection matrix appearing in the R.H.S. 
of (\ref{transv}) in our case is given by,
$\omega_m^{\hat{\mu}\hat{\nu}} = 
{1\over 2} ({\cal E}^T {\cal G}^{-1} {\cal E}_{, m} -
{\cal E}^T_{, m} {\cal G}^{-1} {\cal E})^{\hat{\mu}\hat{\nu}}$
and has a form:
\bea
\omega_m^{\hat{\mu}\hat{\nu}} = {1\over 2} 
\pmatrix{0 & {b_{,m}\over \sqrt{g-a}} & {-b_{,m}\over \sqrt{g-a}}\cr 
{-b_{,m}\over \sqrt{g-a}} & 0 &  {-g_{,m}\over g} + E_{,m} \cr
{b_{,m}\over \sqrt{g-a}} & {g_{,m}\over g}-E_{,m} & 0 }. 
               \label{spinconn}
\eea
Similarly $H_m^{\hat{\mu}\hat{\nu}} \equiv 
({\cal E}^T {\cal G}^{-1} {\cal B}_{,m} {\cal G}^{-1} 
{\cal E})^{\hat{\mu}\hat{\nu}}$ is given by another 
antisymmetric matrix:
\bea
H_m^{\hat{\mu}\hat{\nu}} = \pmatrix{0 & {b_{,m}\over \sqrt{g-a}} &
{-b_{,m}\over \sqrt{g-a}} \cr {-b_{,m}\over \sqrt{g-a}} & 
0 & {-g_{,m}\over g} \cr
{b_{,m}\over \sqrt{g-a}} & {g_{,m}\over g} & 0}. 
                   \label{3form}
\eea
Equation (\ref{transv}) then implies for $\delta \psi_m =0$:
\be
\partial_m \epsilon_L + {E_{,m}\over 4} \Gamma_{\hat{0}\hat{8}}
\epsilon_L = 0,    \label{epsil1} 
\ee
\be
\partial_m \epsilon_R + {1\over 4}[ (- {2g_{,m}\over g} + E_{,m}) 
\Gamma_{\hat{0}\hat{8}} + {2 b_{,m}\over {\sqrt{g-a}}} 
(\Gamma_{\hat{9}\hat{0}} - \Gamma_{\hat{9}\hat{8}})]\epsilon_R = 0.
                       \label{epsir1}
\ee

The variation of the gravitino 
components $\psi_{\mu}$, eqn.(\ref{longi}) can be rewritten as
\be 
\delta \psi_{\mu} \equiv {1\over 4} ({\cal G}^{,\hat{m}}{\cal G}^{-1} 
{\cal E})_{\mu}^{\hat{\nu}}
\Gamma_{\hat{\nu}\hat{m}} \eta 
- {1\over 4} ({\cal B}^{,\hat{m}}{\cal G}^{-1}{\cal E})_{\mu}^{\hat{\nu}}
\Gamma_{\hat{\nu}\hat{m}} \eta^*. \label{longi2}
\ee
To simplify this further we write down the matrices appearing in 
the R.H.S. of this equation: 
\bea
{\cal G}_{,m} {\cal G}^{-1} {\cal E} = 
\pmatrix{0 & {-b_{,m}\over \sqrt{g-a}} & {b_{,m}\over \sqrt{g-a}}\cr
b_{,m} & {{g_{,m} - a_{,m} - b b_{,m}}\over \sqrt{g-a}} &
\sqrt{g-a}{g_{,m}\over g} - {{g_{,m} - a_{,m} - b b_{,m}}\over\sqrt{g-a}} \cr
b_{,m} & -{a_{,m}\over \sqrt{g-a}} - {{b b_{,m}}\over \sqrt{g-a}} & 
\sqrt{g-a}{g_{,m}\over g} + {{a_{,m} + b b_{,m}}\over\sqrt{g-a}}},
              \label{spinconn2}
\eea
and
\bea
{{\cal B}_{,m} {\cal G}^{-1} {\cal E}} = 
\pmatrix{0 & {b_{,m}\over \sqrt{g-a}} & {-b_{,m}\over 
\sqrt{g-a}} \cr
b_{,m} & { {b b_{,m}}\over \sqrt{g-a}} & \sqrt{g-a}{g_{,m}\over g}-
{ {b b_{,m}}\over \sqrt{g-a}} \cr
{b_{,m}} & {{g_{,m}}\over {\sqrt{g-a}}} + { b b_{,m}\over \sqrt{g-a}} & 
\sqrt{g-a}{g_{,m}\over g} -{g_{,m}\over {\sqrt{g-a}}}  
-{ b b_{,m}\over \sqrt{g-a}}}.  \label{3form2}
\eea 
These can be used to show that six equations, 
$\delta\psi_{\mu} =0$, following from the 
real and imaginary components of (\ref{longi2}) 
reduce to only two independent ones with 
$\epsilon_L$ and $\epsilon_R$ satisfying the following 
conditions:
\be
 \left(\Gamma_{\hat{0}\hat{m}} - \Gamma_{\hat{8}\hat{m}}\right) 
 \epsilon_L = 0,  \label{epsil2}
\ee
\be
\left( {2 b_{,m}\over {\sqrt{g-a}}}\Gamma_{\hat{9}\hat{m}}
+ (2{g_{,m}\over g} - E_{,m})\Gamma_{\hat{0}\hat{m}} 
+ E_{,m} \Gamma_{\hat{8}\hat{m}}\right) \epsilon_R = 0.
            \label{epsir2}
\ee

Finally the Killing equations following from the variation of the dilatino 
can be written down in the notations introduced above as:
\be
\delta \lambda = 
\partial_m\phi^{(10)} \gamma^{m} (\epsilon_L - i \epsilon_R) 
- {1\over 2} 
({\cal E}^T {\cal G}^{-1} {\cal B}^{,{\hat{m}}}{\cal G}^{-1}{\cal E}
)^{\hat{\beta}\hat{\gamma}}
\Gamma_{\hat{m}\hat{\beta}\hat{\gamma}} (\epsilon_L + i \epsilon_R) =0,
\ee
and using (\ref{3form}) gives:
\be
 \left( 1 + \Gamma_{\hat{0}\hat{8}} \right)\epsilon_L = 0,
               \label{epsil3}
\ee
\be
\left( - \partial_m \phi^{(10)} + {g_{,m}\over g}\Gamma_{\hat{0}\hat{8}}
 - {{b_{,m}}\over {\sqrt{g-a}}}\Gamma_{\hat{9}\hat{0}} 
 + {{b_{,m}}\over {\sqrt{g-a}}}\Gamma_{\hat{9}\hat{8}} \right)
\epsilon_R = 0.      \label{epsir3}
\ee
The last two expressions can also be written in an alternative form,
using (\ref{gab}):
\be
(\epsilon_L - i \epsilon_R) = 
- [\Gamma_{\hat{0}\hat{8}} + 
tanh{\alpha\over 2} e^{E/2} (\Gamma_{\hat{9}\hat{0}}
- \Gamma_{\hat{9}\hat{8}})] (\epsilon_L + i \epsilon_R),
              \label{complex1}
\ee
which will be useful for discussions later on.

Eqns. (\ref{epsil1}), (\ref{epsil2}) and (\ref{epsil3})
therefore provide complete set of conditions that the 
Killing spinors $\epsilon_L$ have to satisfy. Similarly eqns. 
(\ref{epsir1}), (\ref{epsir2}) and (\ref{epsir3}) 
are the conditions to be satisfied by the Killing spinors
$\epsilon_R$. We also observe that the equations satisfied by 
$\epsilon_L$ are identical to the one for neutral strings. 
That is not surprising as the $O(d-1, 1; d-1, 1)$ transformation,
used to generate solution (\ref{betaeq0}) 
from neutral string solutions, act as identity in this sector. 

The derivation of equations satisfied by the spinors also pass
several consistency checks. First of these, 
as mentioned above, was the reduction of six equations in 
(\ref{longi}) into only two in (\ref{epsil2}) and 
(\ref{epsir2}). Moreover, the dilatino variation equations 
(\ref{epsil3}), (\ref{epsir3}) are also equivalent to these.
To show this, one simply has to multiply (\ref{epsir2})
by $(\Gamma_{\hat{0}} - \Gamma_{\hat{8}})$ from left. 
As a result, one gets a single independent constraint
for $\epsilon_L$, and similarly for $\epsilon_R$. 
The equation involving a derivative on the spinors, 
(\ref{epsil1}) and (\ref{epsir1}) 
are also seen to be consistent with these 
constraint equations. We demonstrate this in section-(2.3)
by obtaining a solution for $\epsilon_L$ and $\epsilon_R$ satisying 
all the equations simultaneously. 
Moreover in section-(2.3) we will also see that the  
final constraints,  (\ref{epsir2}) or 
(\ref{epsir3}), satisfy certain consistency conditions on the 
eigen-values of operators appearing in these equations.

\vskip 1 cm

{\bf (2.3) Killing Spinors}
\vskip .7cm

We now present the solution of the Killing equations for
spinors $\epsilon_L$ and 
$\epsilon_R$. As already stated, $\epsilon_L$ satisfies
the same condition as in the neutral case and 
corresponding solution is also identical:
\be
\epsilon_L = e^{E/4} \epsilon_L^0, 
\ee
where $\epsilon_L^0$ is a constant spinor satisfying, 
\be
 ( 1 + \Gamma_{\hat{0}\hat{8}}) \epsilon_L^0= 0. 
\ee

The form of $\epsilon_R$ is more nontrivial.
This is also obvious from the Killing equations 
(\ref{epsir1}), (\ref{epsir2}) and (\ref{epsir3})
that they satisfy. We already noticed that the two non-derivative
equations (\ref{epsir2}) and (\ref{epsir3}) are in fact 
identical. As a result one finally has
only two equations to solve, namely (\ref{epsir1}) and 
(\ref{epsir2}). Howerver before 
starting to solve these, we first show the self-consistency 
of (\ref{epsir2}) by writing it as:
\be
\left( -1 + ({2 g_{,m}\over {g E_{,m}}} - 1 )\Gamma_{\hat{0}\hat{8}}
+ {2b_{,m}\over {E_{,m} \sqrt{g-a}}} \Gamma_{\hat{9}\hat{8}}
\right)\epsilon_R = 0, 
\ee
and after substituting for $g$ and $b$ from equation (\ref{gab}) as:
\be
\left({{cosh^2{\alpha\over 2}e^{-E} + sinh^2{\alpha\over 2}}
\over {cosh^2{\alpha\over 2}e^{-E} - sinh^2{\alpha\over 2}}}
\Gamma_{\hat{0}\hat{8}} 
- {{2 sinh{\alpha\over 2} cosh{\alpha\over 2} e^{-E/2}}\over
{cosh^2{\alpha\over 2}e^{-E} - sinh^2{\alpha\over 2}}}
\Gamma_{\hat{9}\hat{8}}\right)\epsilon_R = \epsilon_R. 
         \label{constraint1}
\ee
A nontrivial check on our algebra in the previous sub-sections,
as well as about $1/2$ supersymmetry of our solution comes from the fact that 
the particular combination of matrices 
appearing in the L.H.S. of the above 
equation is idempotent, with 
only eigen-values $\pm 1$, as required for the 
validity of the above equation. This can be checked by squaring the LHS of 
(\ref{constraint1}).

The derivative equation (\ref{epsir1}) 
can also be simplified for our backgrounds
using (\ref{gab}), (\ref{epsir3}), and can be written as:
\be
\partial_E {\epsilon}_R = {1\over 2} {cosh^2{\alpha\over 2}e^{-E}
\over 
{(cosh^2{\alpha\over 2}e^{-E} - sinh^2{\alpha\over 2})}}
{\epsilon}_R
- {1\over 4} \Gamma_{\hat{0}\hat{8}} {\epsilon}_R,
          \label{derivative1}
\ee
where we have now changed variable from $r \rightarrow E(r)$. 

Now, to present an explicit solution of the Killing equations: 
(\ref{constraint1}) and (\ref{derivative1}) for 
$\epsilon_R$ we choose a basis for the ten dimensional 
Dirac ($\Gamma$) matrices as in \cite{gsw}:
\be
\Gamma_{\hat{0}} = i \sigma_2 \otimes I_{16}, \>\> 
\Gamma_{\hat{8}} =  \sigma_1\otimes I_{16}, \>\> 
\Gamma_{\hat{9}} = \sigma_3\otimes I_{16}.
\ee
Also, we choose $\epsilon_R \equiv \hat{\epsilon}_R \otimes \chi_0$, 
with $\chi_0$ an unconstrained sixteen-dimensional 
constant spinor and $\hat{\epsilon}_R$
is now a representation of Pauli-matrix algebra. 
Then the final equations to solve are: 
\be
 [- I + \sigma_3 + tanh{\alpha\over 2}e^{E\over 2}
 (\sigma_1 - i \sigma_2) ] \hat{\epsilon}_R = 0
\ee
and 
\be
\partial_E \hat{\epsilon}_R = {1\over 2} {cosh^2{\alpha\over 2}e^{-E}
\over 
{(cosh^2{\alpha\over 2}e^{-E} - sinh^2{\alpha\over 2})}}
\hat{\epsilon}_R
- {1\over 4} \sigma_3 \hat{\epsilon}_R,
\ee
where we have now used eqn.(\ref{epsir3}) instead of (\ref{epsir2}) or
(\ref{constraint1}). 
The final solution for the Killing spinor is:
\bea
\hat{\epsilon}_R = {1\over \sqrt{(cosh^2{\alpha\over 2}e^{-E} - 
sinh^2{\alpha\over 2})}}
\pmatrix{ cosh{\alpha\over 2}e^{-E/ 4} \cr
                       sinh{\alpha\over 2}e^{E/ 4}}. 
            \label{epsifinal}
\eea
This Killing spinor reduces to the one for the 
neutral sting for $\alpha =0$, for which we have
\bea
   \hat{\epsilon}_R \rightarrow
\hat{\epsilon}^N_R = e^{E/4} \pmatrix{1 \cr 0},
                               \label{epsi0}
\eea
and implies in our notations:
\be
   (1- \Gamma_{\hat{0}\hat{8}})\epsilon^N_R = 0. 
                       \label{susy0}
\ee
We have therefore explicitly solved for the Killing spinor and 
shown that a charged macroscopic string solution given in 
equation (\ref{betaeq0}) is $1/2$ supersymmetric. The $1/2$ supersymmetry
comes from the fact that half the components of
$\hat{\epsilon}_R$ are related to the remaining ones as
given in an explicit form in equation (\ref{epsifinal})

We now show that the supersymmetry conditions 
(\ref{constraint1}) and the Killing 
spinors (\ref{epsifinal}) for the charged 
case are related to the neutral ones 
through a Lorentz boost. For this 
we parameterize the coefficients of $(\Gamma_{\hat{0}\hat{8}},
\Gamma_{\hat{9}\hat{8}})$ in equation (\ref{constraint1})
as $(cosh\theta, - sinh\theta)$
respectively and note that for 
$\epsilon_R$ in (\ref{epsifinal}) satisfying this equation,
\be
\epsilon_R^N = (cosh{\theta\over 2} - sinh{\theta\over 2}
\Gamma_{\hat{9}\hat{0}})\epsilon_R,
\ee
with
\be 
cosh{\theta\over 2} = 
{{cosh{\alpha\over 2}e^{-{E/2}}}
\over \sqrt{cosh^2{\alpha\over 2}e^{-E} - sinh^2{\alpha\over 2}}},
\>\>\>
sinh{\theta\over 2} = {{ sinh{\alpha\over 2}}\over
\sqrt{cosh^2{\alpha\over 2}e^{-E} - sinh^2{\alpha\over 2}}},
\ee
reduces to the expression (\ref{epsi0}) and 
satisfies (\ref{susy0}), which is  
also the condition satisfied by the Killing 
spinor for the neutral strings $(\alpha = \beta = 0)$. 
Therefore the $1/2$ supersymmetry condition for a 
charged macroscopic string, namely (\ref{constraint1}), is related to the one
for neutral string by the action of a Lorentz boost on the 
spinor $\epsilon_R$. This is expected, as the solution (\ref{betaeq0}) for the
charged macroscopic string is also generted from the neutral ones
by a Lorentz boost in the right-moving sector. Howerver,
we find it interesting to note 
that the action of this Lorentz transformation on 
the spinors is governed by a coordinate
dependent parameter. Only in the $r\rightarrow \infty$ $(E \rightarrow 0)$
limit, this parameter reduces to the one for a global Lorentz 
transformation. This is similar to the 
phase transformation of spinors 
induced by an $SL(2,Z)$ S-duality transformation 
\cite{ortin}. The transformation of the 
spinors are coordinate dependent under $S$-duality tranformations
as well, although
like $O(d-1, 1; d-1, 1)$ transformations, 
the $SL(2)$'s are themselves global. 
\vskip 1cm

{\bf (2.4) $D < 9$ Solutions}
\vskip .7cm

So far we have restricted ourselves to 
$D=9$. Above analysis generalizes to 
the charged Macroscopic String solutions in $D<9$ 
in a sraightforward manner with only minor modifications.
Using the KK procedure metioned in section-(2.1), 
we can once again decompactify these solutions
to ten dimensions. The resulting ten-dimensional metric 
now has a block-diagonal form:
\be
\hat{G}^{(10)} = \pmatrix{ I_{9-D} & & \cr & {\cal G} & \cr
& & I_{D-2}},   \label{dleq9g}
\ee
with $I_{9-D}$ representing an identity matrix for all the 
internal directions ranging from: $(x^{D+1},...,x^{9})$ and
$I_{D-2}$ represents the tranverse space dimensions of
the string in Cartesian coordinates. Matrix 
${\cal G}$ in eqn. (\ref{dleq9g}) is similar to the one 
in (\ref{calbg}) and is now defined in 
a three dimensional space
with coordinates $(x^{D}, x^0, x^{D-1})$, i.e., by replacing 
in eqn.(\ref{calbg}) the coordinates $(x^9, x^8)$ by $(x^D, x^{D-1})$.
Also, the explicit form of ${\cal G}$ is similar to the one 
in (\ref{calbg}) except $E$ is now a $D-2$ dimensional 
Green's funtion (\ref{green}).
Similarly, the antisymmetric tensor is represented by 
a matrix:  
\be
\hat{\cal B}^{(10)} = \pmatrix{ {0} & & \cr & {\cal B} & \cr 
& & 0}.
\ee
The dilaton remains same as in the D-dimensional theory
and is given by the same expression as in 
(\ref{10phi}) with $E$ modified as in (\ref{green}). 

Due to the block-diagonal form of the backgrounds that we have obtained, 
the supersymmetry analysis is exactly same as previously in this
section. We have therefore shown the $1/2$ supersymmetry of
the $\beta =0$ solution in dimensions $D\leq 9$. 
In next section, we also 
work out the supersymmetry of certain $\alpha \neq 0$, 
$\beta \neq 0$ solutions, 
in order to find a $1/4$ supersymmetric network solution 
of charged macroscopic strings later in section-4.

\section{Supersymmetry of $\alpha, \beta \neq 0$ Solutions}  

In this section we write down the $1/2$ supersymmetry conditions for 
cases: $\alpha = -\beta$ and $\alpha = \beta$ in 
equations (\ref{e13})-(\ref{e19b}). 
Here we only write down the supersymmetry 
conditions which are the analogs of (\ref{complex1}) given earlier.
These conditions will be generalized to a maifestly 
$SL(2, Z)$-covariant form later on. 
Although the full solution of the Killing 
equations can also be obtained as in the last section, we do not
present them here.  
\vskip 1cm

{\bf (3.1) $\alpha = -\beta \neq 0$ Solutions}
\vskip .7cm

Once again we first discuss the solution in $D=9$ and 
then generalize them to the lower dimensional cases. 
The solution  in $D=9$ is now characterized by a metric:
\be
ds^2 = -{1\over {1 + {{C cosh^2\alpha}\over r^5}}} {dt}^2 +
        {1\over {1 + {C\over r^5}}} {(dx^8)}^2 + \sum_{i=1}^{7}
        dx^i dx^i.  \label{al=beg}
\ee
The only non-zero component of the antisymmetric tensor is of the
form
\be
B_{0 8} = - {C cosh\alpha\over 2}\left[{1\over {(r^5 + C)}} 
+ {1\over {(r^5 + C cosh^2\alpha)}}\right].
          \label{al=beb}
\ee
We also have a nontrivial modulus parametrizing the $O(1, 1)$
matrix $M_D$ in eqn.(\ref{e18}):
\be
   \hat{G}_{99} \equiv \hat{g} = {{1+ {C cosh^2\alpha\over r^5}}\over
              {1 + {C\over r^5}}}. 
           \label{hatg}
\ee
The two gauge fields appearing in equations (\ref{decomp}),  
(\ref{aredef}) for $D=9$ are of the form:
\be
\hat{A}^1_t = {{C sinh\alpha cosh\alpha}\over { 2 
(r^5 + C cosh^2\alpha)}}, \>\>\> \hat{A}^1_8 = 0, 
\ee
\be
\hat{A}^2_t = 0, \>\>\>
\hat{A}^2_8 = {-{C sinh\alpha}\over { 2 
(r^5 + C)}}. 
\ee

The supersymmetry property of the above solution is obtained
in the same manner as in section (2.1), after
decompactifying the 9-dimensional backgrounds back to 
ten dimensions. The background fields in ten 
dimensions for $\alpha = -\beta$ case are now represented by
$3\times 3$ matrices analogous to the ones in (\ref{calbg}):
\bea 
 \cal{G} = \pmatrix{ {{1+{C cosh^2\alpha\over r^5}}\over 
{1+ {C\over r^5}}} & {C\over r^5}{{cosh\alpha sinh\alpha}\over
{(1 + {C\over r^5})}} & 0\cr
{C\over r^5}{{cosh\alpha sinh\alpha}\over 
{(1 + {C\over r^5})}} & - {[1- {C sinh^2\alpha\over r^5}]\over
(1+{C\over r^5})} & 0 \cr
0 & 0 & {1\over{(1+{C\over r^5})}}}, \label{abcalg}
\eea
\bea 
 {\cal B} = \pmatrix{0 & 0 & - {C sinh\alpha \over {(r^5 + c)}} \cr
0 & 0 & - {C cosh\alpha \over {(r^5+c)}} \cr
{C sinh\alpha\over{(r^5+c)}} & {C cosh\alpha\over{(r^5 + c)}} & 0}, 
                     \label{abcalb}
\eea
and 
\be 
  \phi^{(10)} = - ln (1+{c\over r^5}). \label{al=bed1}
\ee

The $1/2$ supersymmetry conditions is now 
obtained from the dilatino variation (\ref{dilatino}), 
although other equations are expected to give the same answer as
well. We once again need to 
compute the matrix $H_m^{\hat{\mu}\hat{\nu}}$, the analog of the 
one in eqn.(\ref{3form}). It now has a form:
\be
H_m^{\hat{\mu}\hat{\nu}} = {\partial_m {[1 + {C\over r^5}]}
\over {(1+{C\over r^5})}^2} \times
\pmatrix{ 0 & 0 & - {sinh\alpha\over 
\sqrt{\hat{g}G_{88}}} \cr 
0 & 0 & {1\over {\sqrt{G_{tt}G_{88}}}}{{cosh\alpha}
{(1 + {C\over r^5})}\over {(1 +{C cosh^2\alpha\over r^5})}} \cr
{sinh\alpha\over \sqrt{\hat{g}G_{88}}} & 
- {1\over {\sqrt{G_{tt}G_{88}}}}{{cosh\alpha}
{(1 + {C\over r^5})}\over {(1 +{C cosh^2\alpha\over r^5})}}
& 0}.
\ee
Then after some algebra, the $1/2$ supersymmetry condition
is shown to be:
\be
(\epsilon_L - i \epsilon_R) = \left[ -cosh\alpha 
\sqrt{{1+{C\over r^5}}\over {1+{{C cosh^2\alpha}\over r^5}}}
\Gamma_{\hat{0}\hat{8}} + {sinh\alpha\over \sqrt{1 + {C cosh^2\alpha
\over r^5}}}\Gamma_{\hat{9}\hat{8}}\right]
(\epsilon_L + i \epsilon_R). \label{susy-}
\ee
Once again consistency of this equation is seen by observing that 
the matrix appearing in the RHS of (\ref{susy-})
is idempotent.  

In the present case the $1/2$ supersymmetry of the charged
string, as well as that of the corresponding networks that 
will be discussed in section-4, can be argued in another way
as well. As pointed out earlier,
the solution generating transformations contain
the group of constant coordinate transformations as 
a subgroup.
One can show that $\alpha = -\beta$ solutions belong to this
category. For this we note that the metric and antisymmetric tensors
in the ten-dimensional theory, after decompactification, are 
related to the neutral string solutions as:
\bea
{\cal G} = \Lambda {\cal G}_0 \Lambda^T, \>\>\>
{\cal B} = \Lambda {\cal B}_0 \Lambda^T,
             \label{gg0bb0}
\eea
where ${\cal G}_0$ and ${\cal B}_0$ are the ten-dimensional 
backgrounds for the netutral strings: 
\bea
   {\cal G}_0 = \pmatrix{1 & & \cr
              & - {1\over {(1 + {C\over r^5})}} & \cr
       & & {1\over {(1 + {C\over r^5})}}} \label{calg0},
\eea
\bea
   {\cal B}_0 = \pmatrix{ 0 & 0 & 0 \cr
                       0 & 0 & - {C\over {(r^5 + C)}} \cr
                       0 & {C\over {(r^5 + C)}} & },
                   \label{calb0}
\eea
and 
\bea
\Lambda = \pmatrix{cosh\alpha & sinh\alpha & 0 \cr
        sinh\alpha & cosh\alpha & 0 \cr
        0 & 0 & 1 }. 
\eea

We however like to point out that althought the two solutions
are related by the above transformation, they
are still physically different  
in the compactified theory. The generation of charged solutions
through 
decompactification and constant coordinate transformations are
known, including for many examples of black holes such as 
Reissner-Nordstrom from Scharzschild etc.. The transformations 
(\ref{gg0bb0}) in our case only points out 
that many of the classical properties,
including supersymmetry are identical in two theories. In 
next sub-section we will write down the $1/2$ supersymmetry of 
the charged macrocopic strings for $\alpha = \beta$ case.
These are inequivalent solutions with resepct to the neutral ones 
even in ten dimensions.

The generalization of the supersymmetry condition (\ref{susy-}) to 
$D<9$ is once again stratighforward and follows a similar
path as in section-(2.4). As long as the unit vectors 
$n^{(a)}$ and $p^{(a)}$ in eqns. (\ref{e16}), (\ref{e17})
 are chosen to be along a single internal direction, say $x^D$,  
only modification in (\ref{susy-}) comes in the power of $r$ which is
associated with the Green function in the tranverse directions,
in addition to replacing 
$(\Gamma_{\hat{9}}, \Gamma_{\hat{8}}) \rightarrow
(\Gamma_{\hat{D}}, \Gamma_{\hat{D-1}})$. 
A more interestring case is when we parameterize 
them by angular variables
as $n^{(a)} = p^{(a)} \equiv (cos\omega, sin\omega cos\phi,...)$,
in $(10-D)$-dimensional internal space. 
Then $\Gamma_{\hat{9}}$ in eqn.(\ref{susy-}) is replaced by an 
orthogonal combination of $\Gamma$ matrices in $(10-D)$
internal dimensions: 
$\Gamma_{\hat{9}} \rightarrow \Gamma_{\hat{n}}$. 
We will exploit this property in an eight-dimensional example in 
section-(4.2) to show the existence of network type solutions.

\vskip 1 cm

{\bf (3.2) $\alpha = \beta$ Solutions}
\vskip .6cm

In this case the background metric and antisymmetric
tensors are identical to the one in (\ref{al=beg}). 
The modulus field is
now given by, 
\be
\hat{g} = {{1+ {C \over r^5}}\over
              {1 + {C cosh^2\alpha\over r^5}}}.
           \label{hatg-dual}
\ee
Finally the components of the gauge fields are now:
\be
\hat{A}^1_t = 0,\>\>
\hat{A}^1_8 = {{C sinh\alpha}\over { 2 
(r^5 + C)}},
\ee
\be
\hat{A}^2_t = {-{C sinh\alpha cosh\alpha}\over { 2 
(r^5 + C cosh^2\alpha)}}, \>\>\>\hat{A}^2_8 = 0.
\ee

The ten-dimensional beackgrounds are now represented as: 
\bea
{\cal G} = \pmatrix{\hat{g} & 0 & \tilde{b} \cr
0 & - G_{tt} & 0 \cr
\tilde{b} & 0 & G_{88} 
+ {{\tilde{b}}^2\over {\hat{g}}}}, \label{abcalg2}
\eea
with $G_{tt}$ and $G_{88}$ as in (\ref{al=beg}), 
and $\tilde{b} = 
{{C sinh}\alpha\over {(r^5 + C cosh^2\alpha)}}$. 
Antisymmetric tensor is represented as:
\bea
{\cal B} = {C\over {(r^5 + C cosh^2\alpha)}}\pmatrix{ 0 & 
- sinh\alpha cosh\alpha & 0 \cr
sinh\alpha cosh\alpha & 0 & -cosh\alpha \cr
0 & cosh\alpha & 0}, \label{abcalb2}
\eea
and dilaton is given by the expression: 
\be
\phi^{(10)} = - ln(1+ {C cosh^2\alpha\over r^5}).
       \label{al=bed2}
\ee
The inequivalence of the charged solution with respect to the 
netutral ones can be seen by observing that the 
form of the dilaton in eqn.(\ref{al=bed2}) is now different from that in 
(\ref{al=bed1}). A comparison of $\hat{g}$'s in (\ref{hatg-dual})
and (\ref{hatg}) implies that $\alpha = \beta$ solutions are $T$-dual 
with respect to $\alpha = -\beta$ ones. Property of supercharges 
under $T$-duality has been studied in \cite{hassan94,hassan}. We however 
obtain the $1/2$ supersymmetry condition 
by directly using the background solutions.

The final form of the supersymmetry condition is now:
\be
(\epsilon_L - i \epsilon_R) = - \left(
{1\over cosh\alpha }\sqrt{{1+ {C cosh^2\alpha\over r^5}}
\over {1+{C\over r^5}} } \Gamma_{\hat{0}\hat{8}} 
+ tanh\alpha \sqrt{1\over
{1+{C\over r^5}}} \Gamma_{\hat{9}\hat{0}} 
\right)
(\epsilon_L + i \epsilon_R), \label{susy+1}
\ee
and its self-consistency can again be checked by 
observing that the Matrix in the RHS of (\ref{susy+1}) is 
idempotent.

The extension of this result to $D<9$ is 
again straight-forward. The final result is a replacement of 
($\Gamma_{\hat{9}}$, $\Gamma_{\hat{8}}$) by 
($\Gamma_{\hat{D}}$, 
$\Gamma_{\hat{(D-1)}}$) respectively,
for trivial unit vectors $n^{(a)}$ and $p^{(a)}$'s 
pointing only along $x^D$.  At the same time, 
the power of $r$ is modified 
in this equation appropriately to $r^{D-4}$. 
On the other hand,  
when $n^{(a)} = p^{(a)}$ represent a general rotated unit-vector in 
$(10-D)$-dimensional internal space, 
the supersymmetry condition is also modified by 
replacing $\Gamma_{\hat{D}}$ by $\Gamma_{\hat{n}}$. 

We end this section by implementing these changes 
for the case of $(\alpha = \beta)$
Charged Macroscopic Strings in  $D=8$,
by defining unit vectors: $n^{(2)} = p^{(2)} = 
( cos\omega, sin\omega)$. Then $1/2$ supersymmetry condition
is:
\bea
(\epsilon_L - i \epsilon_R)   & = - \left(
 {1\over cosh\alpha }\sqrt{{1+ {C cosh^2\alpha\over r^4}}
 \over {1+{C\over r^4}}} \Gamma_{\hat{0}\hat{7}} 
  + tanh\alpha \sqrt{1\over
{1+{C\over r^4}}} [ cos\omega \Gamma_{\hat{9}\hat{0}} + 
sin\omega \Gamma_{\hat{8}\hat{0}} ] 
 \right) \times &  \cr
 &  \times (\epsilon_L + i \epsilon_R). &   \label{susy+2}
\eea

\section{SL(2, Z)-Multiplets and Network Solutions}

{\bf (4.1) $(p,q)$ Charged Macroscopic String Solutions}
\vskip .6 cm

The $SL(2,Z)$ multiplets of charged macroscopic strings and their 
supersymmetry properties can be written following
\cite{schwarz95,sen97,ortin}.
The bosonic backgrounds for a general charged 
macroscopic string solution is generated in precisely the same 
manner as in \cite{schwarz95} and can be written down using the 
ten-dimensional solutions that we introduced for our
lower dimensional Charged Macroscopic Strings. 
First, the Einstein metric, defined in ten-dimensions:
\be
  G^E_{M N} = e^{- {\phi^{(10)}}/ 4} G^s_{M N}, \label{einst1}
\ee 
for our $(D=9)$ examples of sections-2 and 3 take a form:
\be
{G}^E = e^{-{\phi^{(10)}/ 4}}\pmatrix{{\cal G} & \cr
                           & I_7}, \label{einst2}
\ee
with ${\cal G}$ and $\phi^{(10)}$'s
given for (i) $\beta =0$ in eqns.
(\ref{10dg},\ref{calbg}) and (\ref{10phi}),
(ii) $\alpha = -\beta$ in eqns. (\ref{abcalg}), (\ref{al=bed1}) and 
(iii) $\alpha = \beta$ in eqns. (\ref{abcalg2}), 
(\ref{al=bed2}) respectively. 
The Einstein metric defined by 
(\ref{einst2}) is invariant under 
the $SL(2, Z)$ transformation. Only modification in these
are in the source terms in the Green function (\ref{green}) to 
make it $SL(2,Z)$ invariant\cite{schwarz95}. 
Nonzero components of the antisymmetric
tensor are given by $3\times 3$ matrices:
\be
({\cal B})^{(i)} = {({\cal M}_0^{-1}})_{i j} q_j {\Delta_q}^{-{1\over 2}}
               ({\cal B}) \label{pqcalb}
\ee
with 
$\Delta_q = q_i ({\cal M}_0^{-1})_{i j} q_j$. 
Components $(i = 1, 2)$ in the above equation correspond to 
the NS-NS and R-R sector fields and $({\cal B})$ is a  
$3\times 3$ matrix  given in 
equations (\ref{calbg}), ({\ref{abcalb}), 
(\ref{abcalb2}) for cases (i), (ii) and (iii) 
listed above. The dilaton for the ten-dimensional 
extension of our $(p,q)$-string
($(p, q)\equiv (q_1, q_2)$ denoted above) 
solution is given by the same expression as in eqn.(20)
of \cite{schwarz95}, 
with $A_q$ replaced by $e^{-\phi^{(10)}}$'s
coming from equations (\ref{10phi}), (\ref{al=bed1}) and 
(\ref{al=bed2}) in 
our three examples. We therefore have the $SL(2, Z)$ covariant 
ten-dimesnional backgrounds for the ten-dimensional extension of our
$D=9$ 
Charged Macroscopic String solution. These can be compactified once
again to $D=9$. The
compactification of type II theories to lower dimensions has 
has been discussed in many papers\cite{berg,pope,luroy} 
and we do not persue it
here. The extension of the results to $D< 9$ solutions is 
straightforward as well. 
We now go on to discuss the supersymmetry properties of these
generalized solutions. 

\vskip 1cm

{\bf (4.2) Supersymmetry}
\vskip .6 cm

The supersymmetry of a $(p,q)$-charged macrscopic 
$D\leq 9$ string solutions
can be examined from the ten-dimensional point of view, 
with type IIB Killing equations as obtained from the supersymmetry
variations written in \cite{schwarz83,hassan}. 
It can be argued that the supersymmetry conditions 
that we have written in 
previous sections will be modified only by a phase factor 
for general $(p,q)$-strings.  This becomes clear when one
writes down the most general variation for the dilatino\cite{hassan}, 
in presence of both NS-NS and R-R backgrounds generated in 
section-(4.1).


For our purpose, we however follow a path presented in \cite{ortin}
for the case of four-dimensional theories with $SL(2,Z)$-duality
symemtries. This argument has been applied to the
case of type IIB $SL(2, Z)$-duality as well\cite{sen97} and
uses the fact that  Killing spinors transform under 
$SL(2, Z)$ by a phase. Explicitly for,
\be
\tau \rightarrow {{a\tau + b}\over {c \tau + d}}
         \label{tautr}
\ee
one has:
\be
(\epsilon_L - i \epsilon_R) \rightarrow e^{{i\over 2}{(c\tau + d)}}
(\epsilon_L - i \epsilon_R).   \label{epsitr}
\ee
In fact, as pointed out in \cite{ortin}, 
the transformation property of the spinors given in 
(\ref{epsitr}) holds for 
Killing spinors in general, including when they are explicitly 
dependent on coordinates, such as $r$ in our case. 
We can now use the above tranformation to generate the 
supersymmetry condition for a general $(p,q)$-string starting from that
for $(1,0)$ ones.

We write down these supersymmetry conditions, 
only at asymptotic infinity, namely in the limit $r \rightarrow \infty$. 
This will be sufficient for our present purpose, 
following a line of study of string networks 
concentrating on the asymptotic properties of spinors\cite{sen97,bhatt}.
Although it is of importance to obtain the full supergravity 
solutions for the networks 
and examine complete supersymmetry 
properties, but we do not address the issue here. 

We also note that the above procedure to generate the supersymmetry 
condition of a $(p, q)$-string, from $(1, 0)$ ones, applies in 
Einstein frame whereas our supersymmetry conditions of sections-2
and 3 are written in the string frame. The translation among these
frames involve redefinitions of fields written explicitly in 
Appendix of \cite{hassan} and involve only dilaton-dependent 
scaling factors, when one restricts to the analysis of 
dilatino supersymmetry variation.  
However since the asymptotic values of the
dilaton in all our examples in previous sections 
turn out to be independent of the parameter
$\alpha$ with $\phi \rightarrow 0$ as $r\rightarrow \infty$, 
identical supersymmetry conditions
hold in Einstein frame as well.  They have explicit forms
for $D=9$ examples as: 
\vskip .5cm

\be 
(i)\>\alpha = -\beta: \>
(\epsilon_L - i \epsilon_R) = e^{- i\Phi(p, q, \tau_0)}
\left[ -cosh\alpha 
\Gamma_{\hat{0}\hat{8}} + {sinh\alpha}\Gamma_{\hat{9}\hat{8}}\right]
(\epsilon_L + i \epsilon_R),
\label{(i)}
\ee
\be
(ii)\> \alpha = \beta: \>\>
(\epsilon_L - i \epsilon_R) = - e^{- i\Phi(p, q, \tau_0)} \left(
{1\over cosh\alpha } \Gamma_{\hat{0}\hat{8}}
+ tanh\alpha 
 \Gamma_{\hat{9}\hat{0}} \right)
(\epsilon_L + i \epsilon_R),
\label{(ii)}
\ee
\be
(iii)\> \beta =0, \alpha \neq 0:\>\>
(\epsilon_L - i \epsilon_R) = - e^{-i\Phi(p, q, \tau_0)}
[\Gamma_{\hat{0}\hat{8}} + tanh{\alpha\over 2}  (\Gamma_{\hat{9}\hat{0}}
- \Gamma_{\hat{9}\hat{8}})] (\epsilon_L + i \epsilon_R),
\label{(iii)}
\ee
with $\Phi$ denoting the phase associated with the complex parameter
$p + q \tau_0$ and the subscript of $\tau$ denotes its asymptotic
value. The value of the phase is 
once again given by the same expression, as for the neutral string,
since the transformations that 
generate them from the charged $(1, 0)$-string 
supersymmetry-condition is 
identical to the one for the neutral ones in \cite{sen97}.

\vskip 1cm

{\bf (4.3) Network Solutions}
\vskip .6cm

To obtain the network solutions, we now start with 
case (i) above and find out if arbitrary number of 
$(p, q)$-strings can be arranged in a manner preserving some 
supersymmetry. 
For this we now generalize (\ref{(i)}) further to accommodate
arbitrary orientation of strings in spatial directions. In particular,
for the string making an angle $\theta$ from $x^8$ axis in an 
$x^8 - x^7$ plane, the supersymmetry condition (\ref{(i)})
modifies into:
\bea
(\epsilon_L - i \epsilon_R) & = exp(- i \Phi(p, q, \tau_0))
\left[ ( -cosh\alpha \Gamma_{\hat{0}} +
{sinh\alpha}\Gamma_{\hat{9}})\times \right. \cr
& (cos\theta \Gamma_{\hat{8}} + sin\theta \Gamma_{\hat{7}})
\left. \right]
(\epsilon_L + i \epsilon_R) \label{Theta}
\eea

The network solution with $1/4$ supersymmetry is then 
found from the above equation by identifying the internal 
and space-time orientations of the strings,
namely $\Phi = \theta$. 
Moreover since the above condition is solved by 
spinors satisfying the following conditions:
\bea
\epsilon_L & = - (cosh\alpha \Gamma_{\hat{0}} -
{sinh\alpha}\Gamma_{\hat{9}} ) \Gamma_{\hat{8}} \epsilon_L, \cr
\epsilon_R & = ( cosh\alpha \Gamma_{\hat{0}} -
{sinh\alpha}\Gamma_{\hat{9}} ) \Gamma_{\hat{8}} \epsilon_R, 
            \label{Fstring1}
\eea 
and 
\be
\epsilon_L  = - (cosh\alpha \Gamma_{\hat{0}} -
{sinh\alpha}\Gamma_{\hat{9}} ) \Gamma_{\hat{7}} \epsilon_R, 
           \label{Dstring1}
\ee
which are independent of the the orientation $\theta$, 
we have the possibility of network solution by arranging 
arbitrarily large number of strings, provided 
charge conservations hold on every 3-string junctions. 

Equations (\ref{Fstring1}) and 
(\ref{Dstring1}) are analogous to the supersymmetry conditions
for the F and D-strings respectively in our case. 
We like to point out that for this example,
the existence of a network solution is already gauranteed 
from its existence in the neutral case. This is because of our
earlier observation that $(\alpha = -\beta)$ 
charged solution is generated from neutral ones 
by a group of constant coordinate transformation. This property 
continues to hold even 
for a $(p,q)$-charged macroscopic string solution, as 
the group of constant 
coordinate transformations commutes with $SL(2, Z)$. 
Above results can be generalized to the lower dimensional 
cases by making appropriate replacements 
already mentioned in section-(3.1)

The network solution and its interpretations are more interesting 
in case (ii), namely for $\alpha = \beta$. First, as 
can be noticed from the
supersymmetry condition, eqn.(\ref{(ii)}), a solution like case (i) 
in $D=9$ does not exist. 
This is because, only the first term in the bracket in the RHS of
eqn. (\ref{(ii)}) can be modified, as in eqn.(\ref{Theta}), 
to include an orientation-depence of the string 
through angle $\theta$. The second term
in the bracket, dependent on $\Gamma_{\hat{9}\hat{0}}$,
namely the ones representing the internal and time coordinates, 
remains unchaged under any spatial rotation 
of string in $x^8-x^7$ plane. As a result, solutions like the ones
in eqns. (\ref{Fstring1}, \ref{Dstring1}) do not work. 

To obtain a network solution in this case, with a unique spinor
satisfying the $(p,q)$ string supersymmetry condition 
for their arbitrary 
orientations, one needs to go down to $D\leq 8$. 
This is done by introducing a parameter associated with rotation in 
internal space, in addition to the angle $\theta$ that the 
string now makes with $x^7$ axis in $x^7-x^6$ spatial plane. 
The eight-dimensional supersymmetry conditions, employing 
internal rotations, was already given in eqn.(\ref{susy+2}). 
A modification of this, for nonzero $\theta$ is given as:
\bea
(\epsilon_L - i \epsilon_R) & = - e^{-i\Phi}\left[ 
{1\over cosh\alpha } 
(cos \theta \Gamma_{\hat{0}\hat{7}} + sin\theta 
\Gamma_{\hat{0}\hat{6}}) \right.
& \cr
& + tanh\alpha 
(cos \omega\Gamma_{\hat{9}\hat{0}} + 
sin\omega \Gamma_{\hat{8}\hat{0}}) 
\left. \right] 
(\epsilon_L + i \epsilon_R). 
\eea

To obtain  $\theta$-independent spinor-projections we now identify
\be
\theta = \Phi = \omega. 
\label{identi}
\ee
This identification allows one to solve eqn. (\ref{identi}) 
for $\epsilon$'s which are $\theta$-independent and 
satisfy projection conditions:
\bea
 - ({1\over cosh\alpha} \Gamma_{\hat{0}\hat{7}} +
 tanh\alpha \Gamma_{\hat{9}\hat{0}}) \epsilon_L 
 & = \epsilon_L, \cr
({1\over cosh\alpha} \Gamma_{\hat{0}\hat{7}} +
 tanh\alpha \Gamma_{\hat{9}\hat{0}}) \epsilon_R 
 & =  \epsilon_R,
           \label{Fstring2}
\eea
and 
\be
 - ({1\over cosh\alpha} \Gamma_{\hat{0}\hat{6}} +
 tanh\alpha \Gamma_{\hat{8}\hat{0}}) \epsilon_L 
  = \epsilon_R. \label{Dstring2}
\ee
The conditions (\ref{Fstring2}) and 
(\ref{Dstring2}) are again the analogs of the 
F-string and D-string supersymmetry conditions for the charged 
macroscopic ($D=8$) strings considered here. 
The identifications (\ref{identi}) imply a coupling between the 
$U(1)$ phase coming from S-dualtiy transformation to the one 
coming from the solution generating transformations. Physically 
this can be interpreted as implying a relationship between 
the gauge-charges with $(p,q)$-charges coming from 2-form 
fields. It will be interesting to analyze the precise 
implications of this relationship on the physical properties of the 
networks.

Finally we comment on the case (i) and other charged macroscopic 
string solutions. It is now evident that the condition 
(\ref{(iii)}) is of a form which does not lead to an obvious 
solution for an 
orientation independent projection condition. This can be
related technically to the fact that in this case one has all three
combination of $\Gamma_{\hat{\mu}\hat{\nu}}$ matrices (in $D=9$) 
appearing in eqn.(\ref{(iii)}), unlike in conditions (i) and 
(ii) where only two of the three combinations appeared, 
allowing above solutions. This is the property of other
$\alpha\neq 0$, $\beta\neq 0$  
solutions as well and may be related to the fact that a general 
left-right asymmetric solution generating transformation acts
differently on $\epsilon_L$ and $\epsilon_R$ and is inconsistent 
with the conditions of having a network solution, as they 
require relationships like (\ref{Dstring1}) and 
(\ref{Dstring2}) between them. 

\section{Conclusions}

In this paper we have obtained supersymmetry 
properties of the charged macrscopic strings. We have also 
shown the existence of  a network solution of charged strings. 
Some of these are completely 
inquivalent with respect to the network of neutral string
solutions. 

In the context of network construction, 
it should be pointed out that our exercise only shows the 
presence of a unique Killing spinor at asymptotic infinity
in the presence of large number of $(p, q)$ strings. We do not
present the spinor at arbitrary space-time point. This 
however requires the knowledge of string network solutions
for the full supergravity which is not completely understood 
even for neutral strings, although progress 
in this direction has been reported\cite{rdevi}. 
More precisely, we notice that the Killing spinor
has a coordinate-dependence given by a covariant expression for the  
Green functions, leading to different spatial dependence
for every $(p,q)$-string. It is hoped that the full Killing spinor
of a supergravity solution for these networks will 
be given by a smooth funtion which will 
properly match on to every string in a network.

It will also be interesting to generalize these to non-planar
networks and possibly to find the applications of such networks
to four-dimensional gauge theories\cite{bergmann,4d}. 
Moreover, one can possibly also analyze the possibility of 
network solutions when strings are compactified on other 
manifolds like $K3$ etc. and be able to obtain a 
realization of various BPS states in string theories in 
this manner.

\section{Acknowledgements}
I would like to thank Sudipta Mukherji for useful 
collaboration at the intital stage of this work as well 
as for other fruitful
discussions. I am  also grateful to M. Alishahiha, I. Antoniadis, 
A. Dabholkar, C. Kounnas, G. Mandal and specially S. Fawad Hassan 
and A. Sen for many helpful discussions and comments.

\vfil
\eject

\end{document}